\documentstyle[12pt]{article}
\begin{document}


\vspace{1cm}

\begin{center}
\Large{ON THE HOPF ALGEBRAS GENERATED \\ BY THE YANG-BAXTER
$R$-MATRICES}
\end{center}

\vspace{.5cm}

\begin{center}
\large{A.A.VLADIMIROV}
\end{center}
\begin{center}
\large{Laboratory of Theoretical Physics, \\
Joint Institute for Nuclear Research, \\
Dubna, Moscow region, 141980, Russia}
\end{center}

\vspace{1cm}

\begin{center}
ABSTRACT
\end{center}

We reformulate the method recently proposed for constructing
quasitriangular Hopf algebras of the quantum-double type from the
$R$-matrices obeying the Yang-Baxter equations. Underlying
algebraic structures of the method are elucidated and an illustration
of its facilities is given. The latter produces an example of a new
quasitriangular Hopf algebra. The corresponding universal $\cal
R$-matrix is presented as a formal power series.

\vspace{6cm}

\_\_\_\_\_\_\_\_\_\_\_\_\_\_\_\_\_\_\_\_\_\_\_\_\_\_\_\_\_\_\_\_\_\_\_\_

E-mail: \ alvladim@theor.jinrc.dubna.su

\pagebreak

{\bf 1}. Following the approaches of refs.~\cite{FRT} and~\cite{Ma} we
proposed in~\cite{Vl} a recipe for constructing quantum doubles
(quasitriangular Hopf algebras of the special type~\cite{Dr,Tj})
associated with invertible solutions of the quantum Yang-Baxter
equations (QYBE).  Let us briefly recall this procedure.

It is known~\cite{FRT} that any invertible solution $R$ of QYBE
\begin{equation}
R_{12}R_{13}R_{23}=R_{23}R_{13}R_{12}               \label{1}
\end{equation}
naturally generates a bialgebra $\cal T$ with generators
$\{1,\,t^i_j\}$ and relations
\begin{equation}
R_{12}T_1T_2=T_2T_1R_{12},\ \ \ \ \Delta (T)=T\otimes T,
 \ \ \ \ \varepsilon (T)={\bf 1}     \label{2}
\end{equation}
($t_{ij}$ form a matrix $T$, \ $\Delta$ is a coproduct
and $\varepsilon$ a counit). We can now define an analogous
bialgebra ${\cal U}=\{1,\,u^i_j\}$ by
\begin{equation}
R_{12}U_1U_2=U_2U_1R_{12},\ \ \ \ \Delta (U)=U\otimes U,
 \ \ \ \ \varepsilon (U)={\bf 1},     \label{3}
\end{equation}
and introduce a pairing between these two,
\begin{equation}
<U_1,T_2>=R_{12}\,,  \label{4}
\end{equation}
as a bilinear map $<\cdot,\cdot>:{\cal U}\otimes {\cal T}
\rightarrow {\cal K}$ into the underlying field ${\cal K}$. The pairing
(\ref{4}) proves to be consistent with the bialgebra structure
\cite{Vl} but is, as a rule, degenerate. Removing the degeneracy by
factoring out so-called null bi-ideals~\cite{Ma} allows us to introduce
antipodes by the relations
\begin{equation}
<S(U_1),\,T_2>=<U_1,\,S^{-1}(T_2)>=R^{-1}_{12}\,,  \label{5}
\end{equation}
and then establish the quantum-double structure on ${\cal T}\otimes
{\cal U}$ using the original Drinfeld recipe~\cite{Dr,Tj}
\begin{equation}
\alpha b=\sum \sum<S(\alpha_{(1)}),b_{(1)}><\alpha_{(3)},b_{(3)}>
b_{(2)}\alpha_{(2)}\,,   \label{6}
\end{equation}
where
\begin{equation}
\Delta^2(\alpha)=\sum \alpha_{(1)}\otimes \alpha_{(2)}\otimes
\alpha_{(3)},\ \ \ \Delta^2(b)=\sum b_{(1)}\otimes b_{(2)}
\otimes b_{(3)}\,.   \label{7}
\end{equation}
In the case (\ref{2})-(\ref{4}) this recipe results in the well known
formula
\begin{equation}
R_{12}U_1T_2=T_2U_1R_{12}\,. \label{8}
\end{equation}
However, it is not very well known that (\ref{8}) can be interpreted
{}~\cite{Vl} as the quantum-double cross-multiplication condition as
well.

In the present paper we develop the method \cite{Vl} along the
following aspects. Firstly, we change the order of certain steps
described above: a definition of antipode will now precede
the bracketing procedure. This will immediately produce the
($R$-generated) Hopf algebra, because Reshetikhin's result~\cite{Re}
enables one to introduce invertible antipodes explicitly and so give up
implicit definitions (\ref{5}), where the invertibility of $S$ was not
guaranteed. However, after the removal of degeneracy of $<\cdot,\cdot>$
by abovementioned factorization, these two ways lead us to the same
quantum double.

Secondly, we now understand why the cross-multiplication relation
(\ref{8}) appears in its final form actually before (and independently
of) any factorization. We show that ${\cal T}\otimes {\cal U}$ can be
provided with the bialgebra (or the Hopf algebra) structure merely due
to appropriate features of the pairing, though degenerate.

Therefore, in the present version of the method, the quotienting by
null bi-ideals does not look so unpredictably dangerous as it does in
{}~\cite{Ma} and~\cite{Vl}. Now it can at most trivialize the whole
output. To show that sometimes it does not, we perform the construction
of the quantum double for one of the $4\times4\ R$-matrices listed in
{}~\cite{Hi}. The resulting Hopf algebra is by no means trivial and
appears to be quasitriangular. We assume the corresponding universal
$\cal R$-matrix to be a formal power series and evaluate its terms up
to the fourth order.  \vspace{.3cm}

{\bf 2}. Here we are to explain how an antipode can be
introduced \cite{Re} into the $R$-generated bialgebra ${\cal T}$. For
generality, let us consider its inhomogeneous version \cite{Vl}
(cf.~\cite{Be,SWW,Lu}) with generators $\{1,t^i_j,E_p\}$
(we prefer to display all the indices):
\begin{equation}
R^{ij}_{mn}\,t^m_p\,t^n_q=R^{mn}_{pq}\,t^j_n\,t^i_m\,,\ \ \
E_p\,t^j_q=R^{mn}_{pq}\,t^j_n\,E_m\,,  \label{9}
\end{equation}
\begin{equation}
\Delta(t^i_j)=t^i_k\otimes t^k_j\,,\ \ \varepsilon(t^i_j)=\delta^i_j\,,
\ \ \Delta(E_j)=E_i\otimes t^i_j+1\otimes E_j\,,\ \ \varepsilon(E_j)=0
\,.  \label{10} \end{equation}
$R$-matrix is a solution of QYBE (\ref{1}):
\begin{equation}
R^{ij}_{lm}R^{lk}_{pn}R^{mn}_{qr}=R^{jk}_{lm}R^{im}_{nr}R^{nl}_{pq}\,.
 \label{11}  \end{equation}

Now let us extend this bialgebra by the inverse elements
$\overline{t}{}^i_j$ (overlining a quantity will always mean its
inverse):
\begin{equation}
t^i_k\,\overline{t}{}^k_j=\overline{t}{}^i_k\,t^k_j=\delta ^i_j\,.
\label{12} \end{equation} As a consequence, the following relations are
to be added to (\ref{9}), (\ref{10}):  \begin{equation}
R^{ij}_{mn}\,\overline{t}{}^n_p\,\overline{t}{}^m_q
=R^{mn}_{qp}\,\overline{t}{}^i_m\,\overline{t}{}^j_n\,,\ \ \
R^{im}_{nq}\,\overline{t}{}^j_m\,t^n_p
=R^{mj}_{pn}\,t^i_m\,\overline{t}{}^n_q\,,\ \ \
\overline{t}{}^j_q\,E_p=R^{mj}_{pn}\,E_m\,\overline{t}{}^n_q\,,
\label{13} \end{equation}
\begin{equation}
\Delta(\overline{t}{}^i_j)=\overline{t}{}^k_j\otimes
\overline{t}{}^i_k\,,
\ \ \ \ \ \ \ \varepsilon (\overline{t}{}^i_j)=\delta ^i_j\,.
\label{14} \end{equation}

Further, assume that $R$ admits not only an inverse matrix
$\overline{R}$,
\begin{equation}
R^{ij}_{mn}\overline{R}{}^{mn}_{pq}=\overline{R}{}^{ij}_{mn}R^{mn}_{pq}
=\delta ^i_p\,\delta ^j_q\,, \label{15}
\end{equation}
but also the matrices $\widetilde{R}$ and $\widetilde{\overline{R}}$
(`twisted inverses' for $R$ and $\overline{R}$, respectively):
\begin{equation}
R^{mj}_{pn}\widetilde{R}^{in}_{mq}=\widetilde{R}^{mj}_{pn}R^{in}_{mq}
=\delta ^i_p\,\delta ^j_q\,,\ \ \
\overline{R}{}^{mj}_{pn}\widetilde{\overline{R}}{}^{in}_{mq}
=\widetilde{\overline{R}}{}^{mj}_{pn}\overline{R}{}^{in}_{mq}
=\delta ^i_p\,\delta ^j_q\,.  \label{16}
\end{equation}
Let us define the tensors
\begin{equation}
\Omega ^i_j\equiv \widetilde{R}^{mi}_{jm}\,,\ \ \ \ \ \
\overline{\Omega}{}^i_j\equiv \widetilde{\overline{R}}{}^{mi}_{jm}\,,
\label{17}  \end{equation}
which are inverse to each other,
\begin{equation}
\overline{\Omega }{}^i_k\,\Omega^k_j=\delta ^i_j\,. \label{18}
\end{equation}
This can be easily seen from
$$ \widetilde{R}^{jk}_{ls}\widetilde{\overline{R}}{}^{il}_{tq}=
\overline{R}{}^{pr}_{ts}R^{ik}_{lm}\widetilde{\overline{R}}{}^{lj}_{pn}
\widetilde{R}^{nm}_{qr}\,, $$
which, in turn, is a direct consequence of QYBE (\ref{11}).

In terms of $\Omega $ and $\overline{\Omega}$ one can define~\cite{Re}
an antipode
\begin{equation}
S(t^i_j)=\overline{t}{}^i_j\,,\ \ \ S(\overline{t}{}^i_j)=
\Omega^i_mt^m_n\,
\overline{\Omega}{}^n_j\,,\ \ \ S(E_i)=-E_j\,\overline{t}{}^j_i
\label{19} \end{equation}
and its inverse
\begin{equation}
\overline{S}(t^i_j)=\overline{\Omega}{}^i_m\,\overline{t}{}^m_n\,
\Omega^n_j\,,\ \ \ \overline{S}(\overline{t}{}^i_j)=t^i_j\,,\ \ \
\overline{S}(E_i)=-\overline{S}(t^j_i)E_j\,. \label{20}
\end{equation}
To confirm the correctness of this definition one can use the following
relations:
\begin{equation}
\Omega^n_m\,t^m_j\,\overline{t}{}^i_n=\Omega^i_j\,,\ \ \ \ \
\overline{\Omega}{}^n_m\,\overline{t}{}^m_j\,t^i_n
=\overline{\Omega}{}^i_j\,. \label{21}  \end{equation}
For example,
$$ S(t^i_k\,\overline{t}{}^k_j)=S(\overline{t}{}^k_j)S(t^i_k)
=\Omega^k_m\,
t^m_n\,\overline{\Omega}{}^n_j\,\overline{t}{}^i_k
=\overline{\Omega}{}^n_j\,\Omega^i_n=\delta ^i_j\,.  $$
In its turn, (\ref{21}) is deduced from
\begin{equation}
\widetilde{R}^{mi}_{jn}\,\overline{t}{}^n_q\,t^p_m
=\widetilde{R}^{pn}_{mq}\,t^m_j\,\overline{t}{}^i_n\,, \label{22}
\end{equation}
which is equivalent to the second equality in (\ref{13}).

Thus, we have completed the construction of the $R$-generated Hopf
algebra ${\cal T}$.

Introduce now a similar Hopf algebra, ${\cal U}$, whose generators
$\{1,u^i_j,\overline{u}{}^m_n,F^q\}$ obey the relations
\begin{equation}
u^i_k\,\overline{u}{}^k_j=\overline{u}{}^i_k\,u^k_j=\delta ^i_j\,,
\label{23} \end{equation} \begin{equation}
R^{ij}_{mn}\,u^m_p\,u^n_q=R^{mn}_{pq}\,u^j_n\,u^i_m\,,\ \ \
R^{ij}_{mn}\,\overline{u}{}^n_p\,\overline{u}{}^m_q
=R^{mn}_{qp}\,\overline{u}{}^i_m\,\overline{u}{}^j_n\,,\ \ \
R^{im}_{nq}\,\overline{u}{}^j_m\,u^n_p
=R^{mj}_{pn}\,u^i_m\,\overline{u}{}^n_q\,, \label{24}
\end{equation}
\begin{equation}
F^i\,u^j_p=R^{ji}_{mn}\,u^m_p\,F^n\,,\ \ \ \
F^j\,\overline{u}{}^i_p=\overline{R}{}^{mj}_{pn}\,\overline{u}{}^i_m
\,F^n\,, \label{25}  \end{equation}
\begin{equation}
\Delta(u^i_j)=u^i_k\otimes u^k_j\,,\ \ \
\Delta(\overline{u}{}^i_j)=\overline{u}{}^k_j\otimes\overline{u}{}^i_k
\,,\ \ \ \Delta(F^i)=F^i\otimes 1+u^i_j\otimes F^j\,, \label{26}
\end{equation}
\begin{equation}
\varepsilon (u^i_j)=\varepsilon (\overline{u}{}^i_j)=\delta ^i_j\,,
\ \ \ \ \ \ \varepsilon (F^i)=0\,, \label{27}
\end{equation}
\begin{equation}
S(u^i_j)=\overline{u}{}^i_j\,,\ \ \
S(\overline{u}{}^i_j)=\Omega^i_mu^m_n \,\overline{\Omega}{}^n_j\,,\ \ \
S(F^i)=-\overline{u}{}^i_jF^j\,, \label{28} \end{equation}
\begin{equation}
\overline{S}(u^i_j)=\overline{\Omega}{}^i_m\overline{u}{}^m_n
\,\Omega^n_j\,,\ \ \ \overline{S}(\overline{u}{}^i_j)=u^i_j\,,\ \ \
\overline{S}(F^i)=-F^j\overline{S}(u^i_j)\,. \label{29}
\end{equation}

Now we can define a pairing~\cite{Vl} $<\cdot,\cdot>:{\cal U}\otimes
{\cal T}\rightarrow {\cal K}$ as follows (all nonzero brackets of the
generators are listed):
\begin{equation}
<u^i_j,t^p_q\,>=<\overline{u}{}^i_j,\overline{t}{}^p_q\,>=R^{ip}_{jq}
\,,\ \ <u^i_j,\overline{t}{}^p_q\,>=\widetilde{R}^{ip}_{jq}\,,\ \
<\overline{u}{}^i_j,t^p_q\,>=\overline{R}{}^{ip}_{jq}, \label{30}
\end{equation}
\begin{equation}
<u^i_j,1>=<\overline{u}{}^i_j,1>=<1,t^i_j>=
<1,\overline{t}{}^i_j>=<F^i,E_j>=\delta^i_j\,. \label{31}
\end{equation}
This pairing is of the antidual type, i.e. the conditions
$$ <\alpha\beta,a>=<\alpha\otimes\beta,\Delta(a)>\,,\ \ \
<\Delta(\alpha),a\otimes b>=<\alpha,ba>\,, $$
\begin{equation}
\varepsilon(a)=<1,a>\,,\ \ \ \ \ \varepsilon(\alpha)=<\alpha,1>\,,
\label{32} \end{equation}
$$  <S(\alpha),a>=<\alpha,\overline{S}(a)>\,,\ \ \
<\overline{S}(\alpha),a>=<\alpha,S(a)> $$
are fulfilled. The proof is straightforward~\cite{Vl} (cf.~\cite{Ma}).

Note that the relation (\ref{5}) is recovered as well:
$$ <S(u^i_j),t^p_q\,>=<\overline{u}{}^i_j,t^p_q\,>
=\overline{R}{}^{ip}_{jq}\equiv (R^{-1})^{ip}_{jq}\,. $$
However, in the present approach, unlike~\cite{Ma,Vl}, an antipode is
defined in an explicit way and is invertible by construction.
\vspace{.3cm}

{\bf 3}. Now we are in a position to transform ${\cal T}\otimes\cal U$
into a quantum double. To achieve this, one has to remove the degeneracy
of the pairing (\ref{30}), (\ref{31}). This can be done~\cite{Ma} by
factoring out null bi-ideals in ${\cal T}$ and ${\cal U}$ (the
procedure is of course consistent with their Hopf algebra structure as
well). After the factorization, ${\cal T}$ and ${\cal U}$ become the
antidual pair of Hopf algebras, so the recipe (\ref{6}) can be applied
to produce the cross-multiplication rules peculiar to the quantum
double. They are:
$$ R^{ij}_{mn}\,u^m_p\,t^n_q=R^{mn}_{pq}\,t^j_n\,u^i_m \,,\ \
R^{ij}_{mn}\,\overline{t}{}^n_q\,\overline{u}{}^m_p
=R^{mn}_{pq}\,\overline{u}{}^i_m\,\overline{t}{}^j_n\,,\ \
R^{im}_{nq}\,\overline{t}{}^j_m\,u^n_p=
R^{mj}_{pn}\,u^i_m\,\overline{t}{}^n_q\,, $$
\begin{equation}
\overline{R}{}^{mj}_{pn}\,\overline{u}{}^i_m\,t^n_q
=\overline{R}{}^{in}_{mq}\,t^j_n\,\overline{u}{}^m_p \,,\ \
u^i_p\,E_q=R^{mn}_{pq}\,E_n\,u^i_m\,,\ \
\overline{u}{}^i_p\,E_q=\overline{R}{}^{in}_{mq}\,E_n\,
\overline{u}{}^m_p\,,
\label{33}
\end{equation}
$$ t^i_p\,F^j=R^{ji}_{mn}\,F^m\,t^n_p\,,\ \
F^i\overline{t}{}^j_q=R^{in}_{mq}\overline{t}{}^j_nF^m\,,\ \
E_jF^i-F^iE_j=t^i_j-u^i_j\,. $$

An interesting fact here is that the role of the factorization
procedure seems to be not so great: it only ensures antiduality
(non-degenerate pairing) but does not affect the explicit form of the
relations (\ref{33}). Really, the latter is determined entirely by the
recipe (\ref{6}) prior to any factorization. So the
cross-multiplication relations of the quantum double take their right
form even if there is no quantum double!

An explanation of this puzzle is the following: the
cross-multiplication structure (\ref{6}) on ${\cal T}\otimes{\cal U}$
is not characteristic of the quantum double only. It occurs quite
naturally if certain conditions (weaker than the quantum-double ones)
are satisfied.  To show this is the aim of the following two
Propositions.

{\bf Proposition 1}. Let ${\cal A}$ and ${\cal B}$ be bialgebras and
let there exist two pairings,\\ $<\cdot,\cdot>:{\cal B}\otimes {\cal A}
\rightarrow {\cal K}$ and $<\!<\cdot,\cdot>\!>:{\cal B}\otimes {\cal A}
\rightarrow {\cal K}$, with the antidual-type properties
$$ <\alpha\beta,a>=<\alpha\otimes\beta,\Delta(a)>\,,\ \ \
<\Delta(\alpha),a\otimes b>=<\alpha,ba>\,, $$
\begin{equation}
<\!<\alpha\beta,a>\!>=<\!<\beta\otimes\alpha,\Delta(a)>\!>\,,\ \ \
<\!<\Delta(\alpha),a\otimes b>\!>=<\!<\alpha,ab>\!>\,, \label{34}
\end{equation}
$$ <\alpha ,1>=<\!<\alpha ,1>\!>=\varepsilon (\alpha )\,,\ \ \
<\tilde{1},a>=<\!<\tilde{1},a>\!>=\varepsilon (a)\,, $$
and an additional relation
\begin{equation}
<\!<_1<_2\Delta(\alpha ),\Delta(a)>\!>_1>_2=
<_1<\!<_2\Delta(\alpha ),\Delta(a)>_1>\!>_2=
\varepsilon (\alpha )\varepsilon (a)\,. \label{35}
\end{equation}
Then the cross-multiplication rule (cf. (\ref{6}))
\begin{equation}
\alpha b=\sum \sum<\!<\alpha_{(1)},b_{(1)}>\!><\alpha_{(3)},b_{(3)}>
b_{(2)}\alpha_{(2)}   \label{36}
\end{equation}
establishes the bialgebra structure on ${\cal A}\otimes {\cal B}$.

In (\ref{34}) $\tilde{1}$ is the unit of $\cal B$, and $<\!<_1<_2$ in
(\ref{35}) indicates that $<\!<\cdot,\cdot>\!>$-operation deals with the
left multipliers in tensor products; whereas $<\cdot,\cdot>$ with the
right ones.

{\bf Proof}. Fix the bases $\{e_i\}$ in $\cal A$ and $\{e^j\}$ in $\cal
B$. Denoting the structure constants and the pairing tensors (which are
in general degenerate) as follows,
$$ e_i\,e_j=c^k_{ij}\,e_k,\ \  \Delta(e_i)=f_i^{jk}(e_j\otimes e_k),\ \
\varepsilon (e_i)=\varepsilon _i\,,\ \ 1=E^ie_i\,, $$
\begin{equation}
e^ie^j=\tilde{f}{}_k^{ij}e^k\,,\ \ \Delta(e^i)=\tilde{c}{}^i_{jk}
(e^k\otimes e^j)\,,\ \ \varepsilon (e^i)=\tilde{E}^i\,,\ \
\tilde{1}=\tilde{\varepsilon}_i\,e^i\,, \label{37}
\end{equation}
$$ <e^i,e_j\,>=\eta ^i_j\,,\ \ \ \ \
<\!<e^i,e_j\,>\!>=\chi ^i_j\,, $$
we may list the relations between them which are caused by the
bialgebra structure of $\cal A$ and $\cal B$,
$$ c^k_{ij}\,c^m_{kn}=c^m_{ik}\,c^k_{jn}\,,\ \
c^k_{ij}E^j=c^k_{ji}E^j=\delta ^k_i\,,\ \
c^k_{ij}\,\varepsilon _k=\varepsilon _i\,\varepsilon _j\,, $$
\begin{equation}
f_n^{ij}f_m^{nk}=f_m^{in}f_n^{jk}\,,\ \
f_i^{jk}\varepsilon _k=f_i^{kj}\varepsilon _k=\delta ^j_i\,,\ \
f_i^{jk}E^i=E^jE^k\,, \label{38}
\end{equation}
$$ c^k_{ij}\,f_k^{rs}=f_i^{mn}f_j^{pq}\,c^r_{mp}\,c^s_{nq}\,,\ \
E^i\varepsilon _i=1 $$
(the same for quantities with a tilde), by the properties (\ref{34}),
\begin{equation}
\eta ^m_k\tilde{f}{}_m^{ij}=\eta ^i_m\eta ^j_nf_k^{mn}\,,\
\eta ^i_mc^m_{jk}=\eta ^m_j\eta ^n_k\tilde{c}{}^i_{mn}\,,\
\chi ^m_k\tilde{f}{}_m^{ij}=\chi ^i_m\chi ^j_nf_k^{nm}\,,\
\chi ^i_mc^m_{jk}=\chi ^m_j\chi ^n_k\tilde{c}{}^i_{nm}\,,\label{39}
\end{equation}
\begin{equation}
\eta ^i_jE^j=\chi ^i_jE^j=\tilde{E}^i\,,\ \ \ \ \
\eta ^i_j\,\tilde{\varepsilon}_i=\chi ^i_j\,\tilde{\varepsilon}_i=
\varepsilon _j\,,  \label{40} \end{equation}
and by the relation (\ref{35}),
\begin{equation}
\tilde{c}{}^i_{mn}\eta ^m_q\chi ^n_pf_j^{pq}=
\tilde{c}{}^i_{mn}\eta ^n_p\chi ^m_qf_j^{pq}=
\tilde{E}^i\varepsilon _j\,. \label{41}
\end{equation}

Now (\ref{36}) reads
\begin{equation}
e^ie_j={\cal P}_{jq}^{ip}\,e_pe^q\,, \ \ \ \ \ \ \ \ {\cal P}_{jq}^{ip}
\equiv \eta ^m_n\,\tilde{c}{}^t_{mq}\,\tilde{c}{}^i_{ts}\,\chi_r^s
\,f_j^{rl}f_l^{pn}\,. \label{42}
\end{equation}
To be convinced that this makes ${\cal A}\otimes {\cal B}$ a bialgebra,
we should verify that the transition from, say, $e^ie_je_k$ to $e_te^n$
can be equally well performed in two different ways, which requires
\begin{equation}
{\cal P}^{ip}_{jq}\,{\cal P}^{qm}_{kn}c^t_{pm}=c^p_{jk}\,{\cal
P}^{it}_{pn} \,. \label{43} \end{equation} Analogously,
$e^ie^je_k\rightarrow e_te^n$ implies \begin{equation} {\cal
P}^{jp}_{kq}\,{\cal P}^{it}_{pm}\,\tilde{f}_n^{mq}=
\tilde{f}_p^{ij}\,{\cal P}^{pt}_{kn}\,, \label{44} \end{equation} and,
at last, $\Delta(e^ie_j)=\Delta(e^i)\Delta(e_j)$ means \begin{equation}
\tilde{c}{}^i_{mn}\,f_j^{pq}\,{\cal P}^{nk}_{pl}\,{\cal P}^{mr}_{qs}=
{\cal P}^{ip}_{jq}\,\tilde{c}{}^q_{sl}\,f_p^{kr}\,. \label{45}
\end{equation}
All the conditions (\ref{43})-(\ref{45}) are verified by direct, though
tedious, calculations with repeated use of (\ref{38})--(\ref{41}). For
example, when proving (\ref{43}) or (\ref{44}), the $cf=ffcc$ relation
from (\ref{38}) is applied twice and $cc=cc$ (or $ff=ff$) many times,
whereas in the case of (\ref{45}) the key property is (\ref{41})
accompanied by numerous applications of $cc=cc$ and $ff=ff$.

A minor problem is caused by checking the conditions
\begin{equation}
E^j{\cal P}^{ip}_{jq}=E^p\delta ^i_q\,,\ \ \
\tilde{\varepsilon}_i{\cal P}^{ip}_{jq}=\tilde{\varepsilon}_q
\,\delta^p_j\,,\ \ \ \varepsilon _p\,\tilde{E}^q\,{\cal P}^{ip}_{jq}=
\tilde{E}^i\varepsilon _j\,, \label{46}
\end{equation}
which reflect the properties of unit and counit. Proposition 1 is
proved.

It is worth noting an alternative form of (\ref{42}),
\begin{equation}
{\cal E}^{mj}_{in}e^i e_j={\cal F}^{mj}_{in}e_j e^i\,, \label{47}
\end{equation}
where
\begin{equation}
{\cal E}^{mj}_{in}=\tilde{c}{}_{ip}^{m}\,\eta ^p_q\,f_n^{qj}\,,\ \ \ \
{\cal F}^{mj}_{in}=\tilde{c}{}_{pi}^{m}\,\eta
^p_q\,f_n^{jq}\,.\label{48} \end{equation} Formula (\ref{47}) is
related to (\ref{42}) through \begin{equation} {\cal
P}_{jq}^{ip}=\overline{\cal E}{}_{mj}^{in}\,{\cal F}_{qn}^{mp}\,,
\label{49}   \end{equation}
with
\begin{equation}
\overline{\cal E}{}^{mj}_{in}=\tilde{c}{}_{ip}^{m}\,\chi^p_q\,
f_n^{qj}\,,\ \ \overline{\cal E}{}_{jn}^{mi}{\cal E}_{ri}^{js}={\cal
E}_{jn}^{mi}\,\overline{\cal E}{}_{ri}^{js}=\delta ^m_r\delta ^s_n\,,
\label{50} \end{equation} and, for completeness, \begin{equation}
\overline{\cal F}{}^{mj}_{in}=\tilde{c}{}_{pi}^{m}\,\chi^p_q\,
f_n^{jq}\,,\ \ \overline{\cal F}{}_{nj}^{im}{\cal F}_{ir}^{sj}={\cal
F}_{nj}^{im}\,\overline{\cal F}{}_{ir}^{sj}=\delta ^m_r\delta ^s_n\,.
\label{51} \end{equation}

The principal goal of the proposition proved was to formulate `minimal'
requirements (\ref{35}) which yet suffice for the cross-multiplication
recipe (\ref{36}) to be fruitful. Properties of the second pairing,
$<\!<\cdot,\cdot>\!>$, as well as (\ref{35}) itself, are motivated by
the anticipation of an antipode. The following proposition states this
explicitly.

{\bf Proposition 2}. Let $\cal A$ and $\cal B$ be the Hopf algebras and
let there exist a pairing $<\cdot,\cdot>:{\cal B}\otimes {\cal A}
\rightarrow {\cal K}$ with the properties (\ref{34}) and, in addition,
\begin{equation}
<S(\alpha),a>=<\alpha,\overline{S}(a)>\,,\ \ \
<\overline{S}(\alpha),a>=<\alpha,S(a)>\,.  \label{52}
\end{equation}
Then the rule (\ref{6}) makes ${\cal A}\otimes {\cal B}$ a Hopf
algebra.

{\bf Proof}. Using the notation
\begin{equation}
S(e_i)=\xi ^j_i e_j\,,\ \ \overline{S}(e_i)=\sigma ^j_i e_j\,,\ \
S(e^j)=\tilde{\sigma}^j_i e^i\,,\ \ \overline{S}(e^j)=
\tilde{\xi}^j_i e^i\,, \label{53}
\end{equation}
we write down the Hopf-algebra properties of $\cal A$ and $\cal B$ as
$$ \sigma ^k_i \xi ^j_k=\xi ^k_i \sigma ^j_k=\delta ^j_i\,,\ \
E^j\xi ^i_j=E^j\sigma ^i_j=E^i\,,\ \
\varepsilon _i \xi ^i_j=\varepsilon _i\sigma ^i_j=\varepsilon _j\,, $$
\begin{equation}
c^k_{ij}\,\xi ^m_k=c^m_{pq}\,\xi ^q_i\xi ^p_j\,,\
c^k_{ij}\,\sigma ^m_k=c^m_{pq}\,\sigma ^q_i\sigma ^p_j\,,\
f_k^{ij}\xi ^k_m=f_m^{pq}\xi ^i_q\,\xi ^j_p\,,\
f_k^{ij}\sigma ^k_m=f_m^{pq}\sigma ^i_q\,\sigma ^j_p\,, \label{54}
\end{equation}
$$ c^j_{nr}\xi ^r_s f_i^{ns}= c^j_{rn}\xi ^r_s f_i^{sn}=
c^j_{nr}\sigma ^r_s f_i^{sn}=c^j_{rn}\sigma^r_sf_i^{ns}=
E^j\varepsilon _i\,, $$
(the same for quantities with a tilde), and the conditions (\ref{52}) as
\begin{equation}
\eta ^k_i\,\tilde{\sigma}^j_k=\eta ^j_k\,\sigma ^k_i\,,\ \ \ \ \ \
\eta ^k_i\tilde{\xi}^j_k=\eta ^j_k\,\xi^k_i\,. \label{55}
\end{equation}
The bialgebra part of the proof is already done in the Proposition 1
because of the following identification:
$$ <\!<\alpha ,a>\!>\equiv <S(\alpha),a>\,, \ \ {\rm i.e.}\ \
\chi ^j_i=\eta ^j_k\sigma ^k_i=\eta ^k_i\tilde{\sigma}^j_k\,. $$
The conditions (\ref{35}) are readily checked,
$$ <\!<_1<_2\Delta(\alpha),\Delta(a)>\!>_1>_2\equiv <(S\otimes id)\circ
\Delta(\alpha),\Delta(a)> $$
$$  =<m\circ(S\otimes id)\circ\Delta(\alpha),a>=\varepsilon (\alpha)
<\tilde{1},a>=\varepsilon (\alpha)\,\varepsilon(a)\,. $$
So it remains to prove that $S(e^ie_j)=S(e_j)S(e^i)$, i.e.
\begin{equation}
{\cal P}_{jq}^{ip}\,\tilde{\sigma}^q_r\,\xi ^s_p\,{\cal P}_{sm}^{rk}
=\xi ^k_j\,\tilde{\sigma}^i_m\,. \label{56}
\end{equation}
It can be done applying the relations from the second line in
(\ref{54}) four times and then, twice, (\ref{41}).

One easily observes that our bialgebras (Hopf algebras) $\cal T$ and
$\cal U$ in Sect.2 fit the above Propositions. This explains the
appearance of the cross-multiplication relations (\ref{8}),(\ref{33})
prior to factorization. The role of the latter is to produce
ortonormalized bases,
$$ <e^i,e_j>\equiv \eta ^i_j=\delta ^i_j\,, $$
that enables one to rewrite (\ref{47}) in the form of
quasicocommutativity condition~\cite{Dr}
\begin{equation}
{\cal R}\Delta(x)=\Delta'(x){\cal R}\,,\ \ \ \ \Delta'\equiv P\circ
\Delta\,,\ \ \ P(a\otimes b)=b\otimes a   \label{57}
\end{equation}
with the universal $\cal R$-matrix
\begin{equation}
{\cal R}=e_i\otimes e^i\,. \label{58}
\end{equation}
\vspace{.3cm}

{\bf 4}. The method described in the present paper creates quantum
doubles out of arbitrary invertible Yang-Baxter $R$-matrices taken as
an input. However, an output (quasitriangular Hopf algebras) might
sometimes appear almost trivial if the factorization involved were
`rude' enough to crash down interesting features of original
bialgebras. Fortunately, this does not necessarily take place.
In~\cite{Vl} (cf.~\cite{Ma}) it is shown how $sl_q(2)$ is recovered by
this method. Another illustration is given below.

Let us take as an input the $R$-matrix~\cite{EOW,Hl,Hi}
\begin{equation} R=\left(
\begin{array}{cccc}1&q&-q&q^2\\0&1&0&q\\0&0&1&-q\\0&0&0&1
\end{array}      \right)  \label{59}
\end{equation}
and consider the homogeneous case of the $R$-generated algebras
(without $E$- and $F$-generators), assuming the notation
$$ T=\left( \begin{array}{cc} a&b\\c&d \end{array} \right)\,,\ \ \ \ \
U=\left( \begin{array}{cc} w&x\\y&z \end{array} \right)\,. $$
To remove the degeneracy of the pairing (\ref{4}), we should require
\begin{equation}
c=y=0\,,\ \ \ ad=da=wz=zw=1\,. \label{60}
\end{equation}
Procedure of Sect.2 results in the Hopf algebra with generators
$\{1,a,\overline{a},b,w,\overline{w},x\}$ whose multiplicative
relations are
$$ [a,b]=q(a^2-1)\,,\ \ [w,x]=q(w^2-1)\,,\ \
[a,x]=qa(w-\overline{w})\,, $$
\begin{equation}  [w,b]=q(a-\overline{a})w\,,\ \ [b,x]=
q(a+\overline{a})x-q(w+\overline{w})b\,,\ \ aw=wa\,, \label{61}
\end{equation}
and the corresponding ones for inverse generators.
We see that $q$ may be absorbed into $b$ and $x$ (so we actually use
the $R$-matrix (\ref{59}) with $q=1$~\cite{DMMZ}). Denoting also
\begin{equation}
a=e^g\,,\ \ \ \ w=e^h\,,\ \ \ \ x=-v\,, \label{62}
\end{equation}
we eventually come to
$$ T=\left( \begin{array}{cc} e^g&b\\0&e^{-g} \end{array} \right)\,,\ \
\ \ \ U=\left( \begin{array}{cc} e^h&-v\\0&e^{-h} \end{array}
\right)\,.  $$
The elements of these matrices form the Hopf algebra
$$ [g,b]=[h,b]=e^g-e^{-g}\,,\ \ \ [g,v]=[h,v]=e^{-h}-e^h\,, $$
$$ [b,v]=(e^g+e^{-g})v+(e^h+e^{-h})b\,,\ \ [g,h]=0\,, $$
\begin{equation}
\Delta(b)=e^g\otimes b+b\otimes e^{-g}\,,\ \ \
\Delta(v)=e^h\otimes v+v\otimes e^{-h}\,, \label{63}
\end{equation}
$$ \Delta(g)=g\otimes 1+1\otimes g\,,\
\Delta(h)=h\otimes 1+1\otimes h\,,\ S^{\pm1}(g)=-g\,,\
S^{\pm1}(h)=-h\,, $$
$$ S^{\pm1}(b)=-b\pm e^g\mp e^{-g}\,,\ \ \
S^{\pm1}(v)=-v\mp e^h\pm e^{-h}\,. $$
The pairing relations are the following:

$$ <1,1>=<h,b>=<v,g>=1\,,\ \ \ <v,b>=-1\,, $$
\begin{equation}
<1,b>=<1,g>=<h,1>=<v,1>=<h,g>=0\,. \label{64}
\end{equation}

By construction, the Hopf algebra (\ref{63}) has to be a quantum
double. In particular, it should possess a universal $\cal R$-matrix.
Assuming exponential Ansatz, we can write down several terms of its
formal power expansion in $g$ and $h$:
\begin{equation}
{\cal R}={\rm exp}\{g\otimes v+b\otimes h-\frac{1}{6}(g\otimes hvh
+gbg\otimes h+g^2\otimes (hv+vh)+(gb+bg)\otimes h^2)+\ldots \}\,,
\label{65}  \end{equation}
where discarded terms are of the fifth order in $g$ and
$h$. To check (\ref{57}) and the quasitriangularity conditions
\begin{equation}
(\Delta\otimes id){\cal R}={\cal R}_{13}{\cal R}_{23}\,,\ \ \
(id\otimes \Delta){\cal R}={\cal R}_{13}{\cal R}_{12} \label{66}
\end{equation}
for the $\cal R$-matrix (\ref{65}), the program FORM ~\cite{Ve} has been
essentially used.

A detailed study of this and other $R$-generated quasitriangular Hopf
algebras is a subject of further investigations.
\vspace{.3cm}

I am grateful to L.Avdeev, A.Isaev and P.Pyatov for stimulating
discussions.

\end{document}